\documentstyle [12pt] {article}

\catcode`@=12
\begin {document}
\thispagestyle{empty}
\begin{flushright} UCRHEP-T96\\July 1992\
\end{flushright}
\vspace{0.5in}
\begin{center}
{\Large \bf Decay of the Z Boson into Scalar Particles\\}
\vspace{0.3in}
\vspace{1.5in}
{\bf T. V. Duong and Ernest Ma\\}
\vspace{0.3in}
{\sl Department of Physics\\}
{\sl University of California\\}
{\sl Riverside, California 92521\\}
\vspace{1.5in}
\end{center}
\begin{abstract}\
In extensions of the standard model, light scalar particles are often possible
because of symmetry considerations.  We study the decay of the Z boson into
such particles.  In particular, we consider for illustration the scalar sector
of a recently proposed model of the 17-keV neutrino which satisfies all
laboratory, astrophysical, and cosmological constraints.
\end{abstract}

\newpage
\section{Introduction}

In the standard model, the Z boson may decay into the Higgs boson H and a
fermion-antifermion pair.  From the absence of such events, it has been
deduced that $m_H > 59~GeV$.\cite{1}  On the other hand, in extensions
of the standard model with a richer scalar sector, there are usually many
more scalar particles, some of which may be light enough to be decay
products of the Z, but with perhaps different signatures.  For example, the
Z boson may now decay into two different scalar particles.  Such a process
is usually unsuppressed if kinematically allowed and serves to set
stringent limits on the masses of the particles involved.  There is however
an important exception.  If a certain scalar particle is coupled to the Z
boson only in association with another heavier particle, then it may in fact
be light and not be produced in a two-body decay.  Three-body decays should
then be considered and could well become a first hint for any new physics
beyond the standard model.

In Section 2 we present a specific example of an extended scalar sector which
goes with a recently proposed model\cite{2} of a possible 17-keV neutrino
which satisfies all laboratory, astrophysical, and cosmological
constraints.\cite{3}
In Section 3 we single out a particular scalar boson of this model and show
how it can be light and still be consistent with all present experimental
data.  In Section 4 we discuss the decay of the Z boson into this scalar
particle as a possible means of discovering its existence.  Finally in
Section 5, there are some concluding remarks.

\section{Example of an Extended Scalar Sector}

To be specific, consider the following extended scalar sector.  Many features
are common to other extensions of the standard model, so it will serve as a
good example of what can be learned from the decay of the Z boson into
scalar particles.  Let
\begin{eqnarray}
V &=& \mu_1^2 \Phi_1^\dagger \Phi_1 + \mu_2^2 \Phi_2^\dagger \Phi_2 +
\mu_{12}^2 (\Phi_1^\dagger \Phi_2 + \Phi_2^\dagger \Phi_1) \nonumber\\
&+& {1 \over 2} \lambda_1 (\Phi_1^\dagger \Phi_1)^2 + {1 \over 2} \lambda_2
(\Phi_2^\dagger \Phi_2)^2 + \lambda_3 (\Phi_1^\dagger \Phi_1)(\Phi_2^\dagger
\Phi_2) + \lambda_4 (\Phi_1^\dagger \Phi_2)(\Phi_2^\dagger \Phi_1) \nonumber\\
&+& m_1^2 \overline \chi_1 \chi_1 + m_2^2 \overline \chi_2 \chi_2 +
m_{12}^2 (\overline \chi_1 \chi_2 + \overline \chi_2 \chi_1) \nonumber\\
&+& {1 \over 2} \eta_1 (\overline \chi_1 \chi_1)^2 + {1 \over 2} \eta_2
(\overline \chi_2 \chi_2)^2 + \eta_3 (\overline \chi_1 \chi_1)(\overline
\chi_2 \chi_2) + \sum_{i,j} f_{ij} (\Phi_i^\dagger \Phi_i)(\overline \chi_j
\chi_j),
\end{eqnarray}
where $\Phi_i = (\phi_i^+,\phi_i^0)$ are doublets and $\chi_i$ are singlets
under the usual $SU(2)~X~U(1)$ electroweak gauge group.  This is the specific
Higgs potential which goes with a recently proposed model\cite{2} of a
possible 17-keV neutrino.  There is an assumed $Z_5$ discrete symmetry with
elements $1,\omega,\omega^2,\omega^{-2}$, and $\omega^{-1}$, with $\omega^5
= 1$, under which the three lepton families, each consisting of a left-handed
doublet $(\nu_i,l_i)_L$ and two right-handed singlets $\nu_{iR}$, $l_{iR}$,
transform as $\omega^{i-1} (i=1,2,3)$.  In addition, $\Phi_1$ transforms as
1, $\Phi_2$ as $\omega^{-2}$, $\chi_1$ as $\omega^2$, and $\chi_2$ as $\omega$.
Furthermore, $\nu_i$ and $l_i$ have lepton number $L=1$ and $\chi_i$ have
$L=-2$.  Note also that the terms $\mu_{12}^2$ and $m_{12}^2$ in $V$ break
$Z_5$ but only softly.

If $\mu_{12}^2=m_{12}^2=0$, then $V$ actually has 4 conserved additive quantum
numbers corresponding to $\Phi_1,\Phi_2,\chi_1,\chi_2$ separately.  As the
vacuum expectation values $<\phi_1^0>,<\phi_2^0>$,\ $<\chi_1>,<\chi_2>$ become
nonzero, 4 Goldstone bosons would appear.  One of them gets "eaten up" by
the Z boson because the gauge symmetry is also broken, but there would
remain 3 physical massless Goldstone bosons.  However, since $\mu_{12}^2$
and $m_{12}^2$ are nonzero, $V$ has only 2 conserved additive quantum
numbers.  The one remaining physical massless Goldstone boson corresponds
to the spontaneous breaking of lepton number and is called the
Majoron.\cite{4}

Let $<\phi_{1,2}^0> = v_{1,2}$ and $<\chi_{1,2}> = u_{1,2}$, then the 4
equations of constraint are
\begin{equation}
\mu_1^2 + \lambda_1 v_1^2 + (\lambda_3 + \lambda_4) v_2^2 + f_{11} u_1^2 +
f_{12} u_2^2 + \mu_{12}^2 v_2/v_1 = 0,
\end{equation}
\begin{equation}
\mu_2^2 + \lambda_2 v_2^2 + (\lambda_3 + \lambda_4) v_1^2 + f_{21} u_1^2 +
f_{22} u_2^2 + \mu_{12}^2 v_1/v_2 = 0,
\end{equation}
\begin{equation}
m_1^2 + \eta_1 u_1^2 + \eta_3 u_2^2 + f_{11} v_1^2 + f_{21} v_2^2 +
m_{12}^2 u_2/u_1 = 0,
\end{equation}
and
\begin{equation}
m_2^2 + \eta_2 u_2^2 + \eta_3 u_1^2 + f_{12} v_1^2 + f_{22} v_2^2 +
m_{12}^2 u_1/u_2 = 0.
\end{equation}
As a result, there are only $12-4=8$ massive degrees of freedom
in the scalar sector.  Two of these correspond to one charged particle
\begin{equation}
h^{\pm} = \sin \beta \phi_1^\pm - \cos \beta \phi_2^\pm
\end{equation}
with mass squared given by
\begin{equation}
m_\pm^2 = -\left( \lambda_4 + {\mu_{12}^2 \over {v_1 v_2}} \right) (v_1^2 +
v_2^2),
\end{equation}
where the angle $\beta$ is defined as
\begin{equation}
\tan \beta \equiv v_2/v_1,
\end{equation}
and
\begin{equation}
(v_1^2 + v_2^2)^{1 \over 2} = (2 \sqrt 2 G_F)^{-{1 \over 2}} \simeq 174~GeV.
\end{equation}
Four others correspond to linear combinations of $Re\phi_{1,2}^0$ and
$Re\chi_{1,2}$ forming a 4 X 4 mass-squared matrix given by
\begin{equation}
{\cal M}^2 = \left( \begin{array}{c@{\quad}c@{\quad}c@{\quad}c}
2\lambda_1v_1^2\!-\!\mu_{12}^2v_2/v_1 & 2(\lambda_3\!+\!\lambda_4)v_1v_2\!+\!
\mu_{12}^2 & 2f_{11}v_1u_1 & 2f_{12}v_1u_2 \\ 2(\lambda_3\!+\!\lambda_4)v_1v_2
\!+\!\mu_{12}^2 & 2\lambda_2v_2^2\!-\!\mu_{12}^2v_1/v_2 & 2f_{21}v_2u_1 &
2f_{22}v_2u_2 \\ 2f_{11}v_1u_1 & 2f_{21}v_2u_1 & 2\eta_1u_1^2\!-\!m_{12}^2
u_2/u_1 & 2\eta_3u_1u_2\!+\!m_{12}^2 \\ 2f_{12}v_1u_2 & 2f_{22}v_2u_2 &
2\eta_3u_1u_2\!+\!m_{12}^2 & 2\eta_2u_2^2\!-\!m_{12}^2u_1/u_2 \end{array}
\right).
\end{equation}
The two remaining massive scalar bosons are
\begin{equation}
h_5^0 = \sqrt 2 (\sin \beta Im \phi_1^0 - \cos \beta Im \phi_2^0)
\end{equation}
and
\begin{equation}
h_6^0 = \sqrt 2 (\sin \gamma Im \chi_1 - \cos \gamma Im \chi_2)
\end{equation}
where $\tan \gamma \equiv u_2/u_1$, with masses squared given by
$-\mu_{12}^2(v_1^2+v_2^2)/v_1v_2$ and $-m_{12}^2(u_1^2+u_2^2)/u_1u_2$
respectively.  This shows explicitly the role of the terms $\mu_{12}^2$
and $m_{12}^2$ in breaking the would-be symmetries $L_e-L_\tau$ and
$L_\mu-L_\tau$.  Of the 4 massless degrees of freedom, 3 are absorbed by
the vector gauge bosons, and one remains as the Majoron
\begin{equation}
\chi_0 = \sqrt {2(u_1^2+u_2^2)} \tan^{-1} \left[ {{\cos \gamma Im \chi_1 +
\sin \gamma Im \chi_2} \over {\cos \gamma Re \chi_1 + \sin \gamma Re \chi_2}}
\right].
\end{equation}

Consider now the Yukawa interactions.  The quarks couple only to $\Phi_1$,
but the leptons interact with all the scalar bosons as follows.\cite{2}
\begin{eqnarray}
{-\cal L} & = & {a \over v_1} (\overline {\nu_1,l_1})_L l_{1R} \left( \!
\begin{array} {c} \phi_1^+ \\ \phi_1^0 \end{array} \! \right) + {b \over v_1}
(\overline {\nu_2,l_2})_L l_{2R} \left( \! \begin{array} {c} \phi_1^+ \\
\phi_1^0 \end{array} \! \right) + {c \over v_1} (\overline {\nu_3,l_3})_L
l_{3R} \left( \! \begin{array} {c} \phi_1^+ \\ \phi_1^0 \end{array} \! \right)
\nonumber \\ & + & {d \over v_2} (\overline {\nu_1,l_1})_L l_{3R} \left( \!
\begin{array} {c} \phi_2^+ \\ \phi_2^0 \end{array} \! \right) + {A \over v_1}
(\overline {\nu_1,l_1})_L \nu_{1R} \left( \! \begin{array} {c} \overline
{\phi_1^0} \\ -\phi_1^- \end{array} \! \right) + {B \over v_1} (\overline
{\nu_2,l_2})_L \nu_{2R} \left( \! \begin{array} {c} \overline {\phi_1^0} \\
-\phi_1^- \end{array} \! \right) \nonumber \\ & + & {C \over v_1} (\overline
{\nu_3,l_3})_L \nu_{3R} \left( \! \begin{array} {c} \overline {\phi_1^0} \\
-\phi_1^- \end{array} \! \right) + {D \over v_2} (\overline {\nu_3,l_3})_L
\nu_{1R} \left( \! \begin{array} {c} \overline {\phi_2^0} \\ -\phi_2^-
\end{array} \! \right) \nonumber \\ & + & {E \over {2u_1}} (\overline
{\nu_2^c}_R \nu_{3R} + \overline {\nu_3^c}_R \nu_{2R}) \chi_1 +
{F \over {2u_2}} \overline {\nu_3^c}_R \nu_{3R} \chi_2 + {\rm H.c.}
\end{eqnarray}
In the above, $l_2$ is exactly $\mu$, but $l_1$ and $l_3$ mix to form $e$
and $\tau$.  Furthermore, $a \simeq m_e$, $b = m_\mu$, $c \simeq m_\tau$,
and $d/c$ is the mixing, which is perhaps of order 0.1.  Let $l_{1L} = c_L e_L
- s_L \tau_L$, $l_{3L} = s_L e_L + c_L \tau_L$, and similarly for $l_{1R}$
and $l_{3R}$, then $s_L \simeq -d/c$, $s_R \simeq s_L m_e/m_\tau$, and
$c_{L,R} \equiv (1-s_{L,R}^2)^{1 \over 2} \simeq 1$.  Since we shall be
interested in the scalar boson $h_5^0$, we note here its interaction with
$e$ and $\tau$.
\begin{eqnarray}
{-\cal L} & = & i m_e \tan \beta (\sqrt 2 G_F)^{1 \over 2} \left( 1 +
{d^2 \over {c^2 \sin^2 \beta}} \right) \overline e \gamma_5 e~h_5^0 \nonumber
\\ & + & i m_\tau \tan \beta (\sqrt 2 G_F)^{1 \over 2} \left( 1 -
{d^2 \over {c^2 \sin^2 \beta}} \right) \overline \tau \gamma_5 \tau~h_5^0
\nonumber \\ & + & {{i m_e (\sqrt 2 G_F)^{1 \over 2}} \over {\sin \beta \cos
\beta}} \left( {d \over c} \right)^3 \overline \tau_L e_R~h_5^0 -
{{i m_\tau (\sqrt 2 G_F)^{1 \over 2}} \over {\sin \beta \cos \beta}}
\left( {d \over c} \right) \overline e_L \tau_R~h_5^0.
\end{eqnarray}

\section{Constraints on the $h_5^0$ Mass}

Consider now the couplings of the W and Z bosons to the various scalar
particles of the previous section.  They are easily extracted from
\begin{equation}
{\cal L} = (D^\mu \Phi_1)^\dagger (D_\mu \Phi_1) + (D^\mu \Phi_2)^\dagger
(D_\mu \Phi_2),
\end{equation}
where the covariant derivative $D_\mu$ is given by
\begin{equation}
D_\mu = \partial_\mu + {{i e \sqrt 2} \over {\sin \theta_W}} (T^+ W_\mu^+ +
T^- W_\mu^-) + {{i e} \over {\sin \theta_W \cos \theta_W}} (T^3 - \sin^2
\theta_W Q) Z_\mu + i e Q A_\mu.
\end{equation}
As a result, it can readily be verified that whereas $Z_\mu Z^\mu
Re \phi_{1,2}^0$ couplings exist, there is no $Z_\mu Z^\mu h_5^0$ coupling.
Hence the experimentally well-studied decay Z $\rightarrow$ "Z" + H, where
the virtual "Z" converts into a fermion-antifermion pair is of no use in
limiting the mass of $h_5^0$.  Furthermore, whereas Z couples to $h^+ h^-$,
it only couples to $h_5^0$ in association with $Re \phi_{1,2}^0$.
Similarly, W only couples to $h_5^0$ in association with $h^\pm$.  Since
these other scalar particles may well be more massive than the Z and W bosons
themselves, there is again no constraint on the mass of $h_5^0$ from their
two-body decays.

The decay $\tau \rightarrow e h_5^0$ is possible if kinematically allowed.
{}From Eq. (15), we find
\begin{equation}
\Gamma = {{\sqrt 2 G_F m_\tau^3} \over {32 \pi \sin^2 \beta \cos^2 \beta}}
\left( {d \over c} \right)^2 \left( 1 - {m^2 \over m_\tau^2} \right)^2,
\end{equation}
where $m$ is the mass of $h_5^0$ and $m_e$ has been neglected.  For $d/c=0.1$
and $\sin^2 \beta = \cos^2 \beta = 0.5$, we have $\Gamma = 3.68 X 10^{-8} (1 -
m^2/m_\tau^2)^2~GeV$, which is some 4 orders of magnitude greater than the
observed $\tau$ decay rate, hence $m > m_\tau$ is required.  Since $h_5^0$
also couples to $\overline {e} \gamma_5 e$, the decay $\tau^\pm \rightarrow
e^\pm e^\pm e^\mp$ is still possible and its amplitude is given by
\begin{equation}
{\cal A} = {{m_e m_\tau G_F} \over {\sqrt 2 \cos^2 \beta}} \left( {d \over c}
\right) \left( 1 + {d^2 \over {c^2 \sin^2 \beta}} \right) {{\overline {e}
(k_2) \gamma_5 e(k_3) \overline {e} (k_1) (1+\gamma_5) \tau (p)} \over
{(p-k_1)^2 - m^2}} - [k_{1,2} \rightarrow k_{2,1}].
\end{equation}
However, it is suppressed by a factor of at least $m_e/m_\tau$ relative to the
usual decay amplitudes of $\tau$, and can thus be safely ignored.  Another
possible decay is $\tau \rightarrow e \gamma$.  Using Eq. (15) we find for
$m > m_\tau$ a branching fraction of less than $10^{-6}$ which is far below
the present experimental upper limit of $2 X 10^{-4}$.\cite{7}  Similarly,
the anomalous magnetic moment of the electron receives from these interactions
a maximum contribution of order $10^{-15}~e/2m_e$, which is negligible
compared to the present experimental error, which is of order
$10^{-11}~e/2m_e$.\cite{7}  Therefore, the only constraint on the mass
of $h_5^0$ from present data remains $m > m_\tau$.

\section{Z Decay into $h_5^0$}

If $h_5^0$ is indeed light, then perhaps we should consider Z $\rightarrow
h_5^0 h_5^0$ as a loop-induced decay.  However, because of angular-momentum
conservation and Bose statistics, a vector particle is absolutely forbidden
to decay into 2 identical scalar particles.  We might try Z $\rightarrow
h_5^0 \chi_0$, which is possible, but its rate is very much suppressed.
Consider now Z $\rightarrow h_5^0 + "Re \phi_{1,2}^0"$, where the virtual
$"Re \phi_{1,2}^0"$ particle converts into a pair of $h_5^0$'s or $\chi_0$'s.
For this possibility, we need to take a look at the triple scalar couplings
contained in Eq. (1).  We find
\begin{eqnarray}
-\cal L & = & v_1 Re \phi_1^0 (\lambda_1 \sin^2 \beta + (\lambda_3 +
\lambda_4) \cos^2 \beta) (h_5^0)^2  \nonumber \\ & + & v_2 Re \phi_2^0
(\lambda_2 \cos^2 \beta + (\lambda_3 + \lambda_4) \sin^2 \beta) (h_5^0)^2
\nonumber \\ & + & u_1 Re \chi_1 (f_{11} \cos^2 \beta + f_{21} \sin^2 \beta)
(h_5^0)^2  \nonumber \\ & + & u_2 Re \chi_2 (f_{12} \cos^2 \beta + f_{22}
\sin^2 \beta) (h_5^0)^2  \nonumber \\ & - & (u_1 Re \chi_1 + u_2 Re \chi_2)
\partial_\mu \chi_0 \partial^\mu \chi_0 / (u_1^2 + u_2^2),
\end{eqnarray}
where the last term actually comes from the $\partial_\mu \overline \chi_i
\partial^\mu \chi_i$ piece of the Lagrangian.  Now Z couples to $h_5^0$ in
association with $\sqrt 2 (\sin \beta Re \phi_1^0 - \cos \beta Re \phi_2^0)$
with coupling strength $e/(2\sin \theta_W \cos \theta_W)$ and $Re \phi_{1,2}^0$
mixes with $Re \chi_{1,2}$ through Eq. (10), hence the decays
\begin{equation}
{\rm Z} \rightarrow h_5^0 \chi_0 \chi_0
\end{equation}
and
\begin{equation}
{\rm Z} \rightarrow h_5^0 h_5^0 h_5^0
\end{equation}
should have nonnegligible rates comparable to that of Z $\rightarrow$ H $f
\overline f$ in the standard model.

Consider first Z $\rightarrow h_5^0 \chi_0 \chi_0$ as shown in Fig. 1.
Assuming that the amplitude is dominated by one scalar intermediate state
of mass $M$, we then have
\begin{equation}
{\cal A} = {{e f_{eff}} \over {\sin \theta_W \cos \theta_W \sqrt {2(u_1^2 +
u_2^2)}}} {{\epsilon \cdot (k_1 - k_2 - k_3)~k_2 \cdot k_3} \over
{(k_2 + k_3)^2 - M^2}}.
\end{equation}
Using $p = k_1 + k_2 + k_3$ and $\epsilon \cdot p = 0$, we find the
spin-averaged amplitude squared in the center of mass to be given by
\begin{equation}
|{\cal A}|_{av}^2 = {{e^2 f_{eff}^2 |\vec {k}_1|^2} \over {6 \sin^2 \theta_W
\cos^2 \theta_W (u_1^2 + u_2^2)}} \left[ 1 + {M^2 \over {(m_Z^2 + m^2 - M^2
- 2 m_Z E_1)}} \right]^2,
\end{equation}
where $m$ is the mass of $h_5^0$, $\vec {k}_1$ its momentum, and $E_1$ its
energy in the rest frame of the Z boson.  Integrating over the invisible
$\chi_0$'s, we obtain the decay energy spectrum
\begin{equation}
{{d\Gamma} \over {dE_1}} = {{|\vec {k}_1|} \over {128 \pi^3 m_Z}}
|{\cal A}|_{av}^2,
\end{equation}
where $E_1$ ranges from $m$ to $(m_Z^2 + m^2)/2m_Z$ and $|\vec {k}_1| =
(E_1^2 - m^2)^{1 \over 2}$.  If $m$ can be neglected, then
\begin{equation}
\Gamma = {{m_Z f_{eff}^2 M^4} \over {384 \pi^3 v^2 u^2}} \left[ - {5 \over
{16z}} + {19 \over 32} - {{7z} \over 24} + {z^2 \over 64} - {{(1\!-\!z)^2
(5\!-\!2z)} \over {16z^2}} \ln (1\!-\!z) \right],
\end{equation}
where $v^2 \equiv v_1^2 + v_2^2$, $u^2 \equiv u_1^2 + u_2^2$, and
$z \equiv m_Z^2/M^2$.  As an illustration, let $f_{eff}^2/4\pi = 0.01$
and $u = v$, then $\Gamma = 3.8 X 10^{-7}~GeV$ for $z = 1$.  Dividing by
the total width\cite{5} $\Gamma_Z = 2.487 \pm 0.010~GeV$, this corresponds
to a branching fraction of about $1.5 X 10^{-7}$, which is an order of
magnitude below the present experimental capability for detection.

Once $h_5^0$ is produced, it will decay into a fermion-antifermion pair.
Hence Z $\rightarrow h_5^0 \chi_0 \chi_0$ has the signature of Z $\rightarrow
f \overline {f} + nothing$, which can also be due to the standard-model
process Z $\rightarrow$ H + "Z", where H $\rightarrow f \overline {f}$ and
"Z" $\rightarrow \nu \overline \nu$.  The big difference is that the only
unknown for the standard-model process is $m_H$, whereas here we have three
unknowns: $m$, $M$, and $f_{eff}$.  However, it is reasonable to assume that
$M$ is of order $m_Z$ and $f_{eff}$ is not much less than unity, so if
$m^2 << m_Z^2$, this process may be observable in the near future at LEP with
more data. To tell it apart from Z $\rightarrow$ H + "Z", we note that the
couplings of $Re \phi_{1,2}^0$ to $f \overline f$ are suppressed by
$m_f/v_{1,2}$ relative to those of the Z boson.  Indeed, $\chi_0 \chi_0$
may well be the dominant decay mode of $Re \phi_{1,2}^0$ so that even if
the latter are produced, their decays would be invisible.\cite{6}

Consider now Z $\rightarrow h_5^0 h_5^0 h_5^0$.  This proceeds as in Fig. 1,
but with $\chi_0$ replaced by $h_5^0$ and the amplitude is the sum of 3 terms
as it must be symmetric with respect to $k_{1,2,3}$.
\begin{equation}
{\cal A} = {{e v \lambda_{eff}} \over {\sin \theta_W \cos \theta_W}} \left[
{{\epsilon \cdot k_1} \over {(p-k_1)^2 - M^2}} + {{\epsilon \cdot k_2} \over
{(p-k_2)^2 - M^2}} + {{\epsilon \cdot k_3} \over {(p-k_3)^2 - M^2}} \right].
\end{equation}
{}From the three-body kinematics, we find that in the center of mass,
\begin{equation}
m < E_1 < {{m_Z^2 - 3m^2} \over {2m_Z}}
\end{equation}
and for a given $E_1$,
\begin{equation}
E_2^{max,min} = {{m_Z - E_1} \over 2} \pm {1 \over 2} |\vec {k}_1| \left(
1 - {{4m^2} \over {m_Z^2 - 2m_Z E_1 + m^2}} \right)^{1 \over 2}.
\end{equation}
This means that $d\Gamma/dE_1$ has a kinematical zero not only at $E_1 = m$
but also at $E_1 = E_1^{max}$ for which $E_2^{max} = E_2^{min} = (m_Z^2 +
3m^2)/4m_Z$.  From Eq. (27) we see also that ${\cal A}$ is zero to order
$M^{-2}$ because $\epsilon \cdot (k_1 + k_2 + k_3) = 0$, hence
we expect in general a significant suppression of the Z $\rightarrow h_5^0
h_5^0 h_5^0$ rate.  Assuming that $m_Z^2 << M^2$, we then have
\begin{equation}
|{\cal A}|_{av}^2 \simeq {{8 \lambda_{eff}^2 m_Z^4} \over {3 M^8}} |E_1 \vec
{k}_1 + E_2 \vec {k}_2 + E_3 \vec {k}_3|^2.
\end{equation}
If we also assume that $m^2 << m_Z^2$, then
\begin{equation}
{{d\Gamma} \over {dE_1}} \simeq {{\lambda_{eff}^2 m_Z^3 E_1^3} \over {48 \pi^3
M^8}} \left( m_Z^2 - {14 \over 3} m_Z E_1 + {28 \over 5} E_1^2 \right).
\end{equation}
This distribution has an interesting shape because there is a local maximum
at $2E_1/m_Z = 0.54$ and a local minimum at $2E_1/m_Z = 0.79$.  This
qualitative feature remains even if we go away from the limit $m^2 << m_Z^2
<< M^2$.  We plot this in Fig. 2 for $m = 10~GeV$ and $M = 100~GeV$.
Note that the kinematical zero at $E_1=E_1^{max}$ for $m \neq 0$ forces
the rising $d\Gamma/dE_1$ to turn over near the end.
Integrating over $E_1$ for $m=0$ and dividing by 6 for the 3 identical
particles in the final state, we find
\begin{equation}
\Gamma \simeq {{m_Z \lambda_{eff}^2} \over {192 \pi^3}} {1 \over 1440}
\left( {m_Z \over M} \right)^8.
\end{equation}
For comparison, let $\lambda_{eff}^2/4\pi = 0.01$ and $M = 2m_Z$, then
$\Gamma \simeq 5.22 X 10^{-9}~GeV$, which shows clearly that unless $M
\approx m_Z$, this rate would be much too small to be of any practical value.

\section{Conclusion}

In summary we have demonstrated how the existence of a light scalar boson can
be consistent with all present experimental data in a specific extension of
the standard model.  It has the potential of being discovered in the future
as a decay product of the Z boson at the level of $10^{-7}$ in branching
fraction.  This result is based on the study of the three-body decays\
Z $\rightarrow h_5^0 \chi_0 \chi_0$ and Z $\rightarrow h_5^0 h_5^0 h_5^0$,
where $h_5^0$ can be as light as the $\tau$ lepton and $\chi_0$ is the
Majoron which is massless.  Although we have used a specific model\cite{2}
for our analysis, such decays of the Z boson are generally present in
extensions of the standard model with two or more scalar doublets and
possibly some singlets.  They may provide the first glimpse of new physics
that is just beyond the reach of present high-energy accelerators.
\vspace{0.3in}
\begin{center} {ACKNOWLEDGEMENT}
\end{center}

This work was supported in part by the U. S. Department of Energy under
Contract No. DE-AT03-87ER40327.

\newpage
\bibliographystyle{unsrt}

\newpage
\begin{center} {FIGURE CAPTIONS}
\end{center}

Fig. 1.  Diagram for the decay Z $\rightarrow h_5^0 \chi_0 \chi_0$.

Fig. 2.  The $\Gamma^{-1}d\Gamma/dE_1$ distribution for Z $\rightarrow
h_5^0 h_5^0 h_5^0$ with $m=10~GeV$ and $M=100~GeV$.

\end{document}